\documentclass[12pt]{iopart}
\usepackage{iopams}  
\usepackage{graphicx,color}

\def\bra#1{\mbox{\boldmath $#1$}^{\mathrm{T}}}
\def\ket#1{\mbox{\boldmath $#1$}}

\begin{document}

\title{Localized eigenvectors of the non-backtracking matrix}

\author{Tatsuro Kawamoto}

\address{Department of Computational Intelligence and Systems Science, 
Tokyo Institute of Technology, 4259-G5-22, Nagatsuta-cho, Midori-ku, Yokohama, Kanagawa, 226-8502, Japan}
\ead{kawamoto@sp.dis.titech.ac.jp}
\vspace{10pt}

\begin{abstract}
In the case of graph partitioning, the emergence of localized eigenvectors can cause the standard spectral method to fail. To overcome this problem, the spectral method using a non-backtracking matrix was proposed. 
Based on numerical experiments on several examples of real networks, it is clear that the non-backtracking matrix does not exhibit localization of eigenvectors. 
However, we show that localized eigenvectors of the non-backtracking matrix can exist outside the spectral band, which may lead to deterioration in the performance of graph partitioning. 
\end{abstract}

%
%
%
%
%

\section{Introduction} \label{Introduction}


Graph partitioning or community detection is a popular topic in the field of complex networks and computer science; the spectral method is a standard heuristic for it. Because the graph partitioning problems are usually formulated as discrete optimization problems, which are NP problems, we require some heuristics. 
In the spectral method, the discreteness constraint is relaxed, and the problem is formulated as a continuous optimization problem. The simplification then reduces to an eigenvalue/eigenvector problem of a matrix (see \cite{Luxburg2007} for detail). 

The commonly used matrices in the literature are the unnormalized Laplacian $L = D - A$, normalized Laplacian $\mathcal{L} = D^{-1/2} L D^{-1/2}$ \cite{Luxburg2007}, and modularity matrix $M = \left( A - \ket{k} \bra{k}/K\right)/K$ \cite{Newman2006,Newman2006PRE}, 
where $A$ is the adjacency matrix; each element represents a node pair and $A_{ij} = 1$ if the node pair $i$ and $j$ are connected by a link, otherwise $A_{ij} = 0$. We only consider undirected graphs and thus $A$ is symmetric.  
The matrix $D$ is a diagonal matrix whose element is equal to the node's degree $k_{i} = \sum_{j} A_{ij}$. 
In the modularity matrix $M$, $\ket{k}$ is a column vector of degrees, $\ket{k} = (k_{1}, \dots, k_{N})^{\mathrm{T}}$ ($\mathrm{T}$ represents the transpose), and $K = \sum_{i=1}^{N} k_{i}$ is the total degree. 
The spectral method utilizes the eigenvectors of large or small eigenvalues; 
for example, in the case of bisection using $L$ and $\mathcal{L}$, the signs of eigenvector elements of the second smallest eigenvalue determines the module assignment of nodes. 
That is, the nodes with the same sign belong to the same module. 
On the other hand, for bisection using $M$, the eigenvector of the largest eigenvalue is referred. 
Though the result of the spectral method obtained by the abovementioned procedure often leads to a good approximation of the true optimum, there is no theoretical guarantee that this is always true. 

In addition to the matrices specified earlier, recently, much attention has been paid to the spectral method using the non-backtracking matrix $B$ \cite{Krzakala2013}, which is defined by the following $2N\times 2N$ matrix: 
\begin{eqnarray}
B = 
\left(\begin{array}{cc}
0 & D- I \\
-I & A\\
\end{array}\right), \label{NBmatrix}
\end{eqnarray}
where $I$ is an identity matrix. 
The matrix $B$ is asymmetric and its spectrum is complex. For graph partitioning, however, one usually focuses on large real eigenvalues. 
The left eigenvector of (\ref{NBmatrix}) has the eigenvector of the form $(\ket{v}^{\mathrm{T}}, -\mu \ket{v}^{\mathrm{T}})$ where $\mu$ is an eigenvalue, and we refer to the sign of $\ket{v}$ of the second-largest eigenvalue for module assignment in the bisection problem. 

The emergence of localized eigenvectors causes the spectral method to fail. 
A localized eigenvector is one whose weight is concentrated around a few components while the remaining components have zero weights or weights close to zero. Furthermore, the localized eigenvector has no information of global community structure. 
For example, if a localized eigenvector is the leading eigenvector of the modularity matrix $M$, the standard bisection method provides a solution that is uncorrelated to the true optimum. 
Indeed, for sparse stochastic block models, localized eigenvectors emerge frequently around the detectability threshold when we use the adjacency matrix $A$ \cite{Krzakala2013} and the Laplacians, $L$ and $\mathcal{L}$ \cite{Kawamoto2015}; this has a considerable effect on the consistency with the planted partition. The details of the stochastic block model are discussed below. 
The spectral partitioning method using the non-backtracking matrix $B$ was originally introduced to avoid localization in such a situation \cite{Krzakala2013}. 

In this paper, first, we briefly review previous works on the localized eigenvector (Sec.~\ref{Review}). We then conduct a comparative analysis of the degree of localization for the abovementioned matrices in several real networks and show that non-backtracking matrix does not exhibit high inverse participation ratios (IPRs) (Sec.~\ref{RealNetworks}). 
Based on this result, it may be assumed that the eigenvectors of the non-backtracking matrix never exhibit localization. However, in Sec.~\ref{LocalizedNB}, we show that localized eigenvectors with real eigenvalues do exist; furthermore, one of them may have the second-largest eigenvalue (Sec.~\ref{MotifDoublingSBM}). 
Finally, some discussion on the localized eigenvectors is given in Sec.~\ref{Discussion}.

\section{Degree of localization}\label{Review}
The degree of localization can be measured by the IPR. 
For an eigenvector $\ket{v}$, the IPR is defined as follows. 
\begin{eqnarray}
\mathrm{IPR} = \frac{ \sum_{i} v_{i}^{4} }{ (\sum_{i} v_{i}^{2})^{2} }. 
\end{eqnarray}
The value of IPR is close to $1$ if $\ket{v}$ is localized and is of $O(N^{-1})$ for a graph with size $N$ if $\ket{v}$ is an extended vector. 
Although localized eigenvectors have been studied for decades, there is still much scope for a more thorough theoretical understanding \cite{Biroli1999,Semerjian2002,Metz2010,Kabashima2012,Kawamoto2015,Nadakuditi2013,Farkas2001,Dorogovtsev2003}. 
The localized eigenvectors of adjacency matrix $A$ in complex networks that exist owing to the hub structure have been frequently discussed in the literature \cite{Goh2001,Farkas2001,Goltsev2012}; in addition, it is known that the localized eigenvectors of the modularity matrix $M$ also exists because of the hub structure \cite{Nadakuditi2013}. 
However, it is noteworthy that this is not always the case, depending on the matrix one considers \cite{PhysRevE.80.026123}. 
Furthermore, there are many researches on the localized eigenvectors in real networks as well \cite{Farkas2001,cucuringu2011localization,Martin2014}. 
To the best of our knowledge, the comparative analysis of localization for commonly used matrices has not been performed thoroughly.

\section{Degree of localization in real networks}\label{RealNetworks}
We investigate the performance of the spectral method of matrices used for graph partitioning in real networks. 
We used the datasets present in \cite{NewmanDataset,SNAP}. 
The networks we consider may not have clear module structures, and for simplicity, we ignore the direction in case of directed networks; further, we deleted self loops if they exist, and converted multiple links to a single link. 

\begin{figure}[!t]
\centering
\includegraphics[width=0.6\columnwidth]{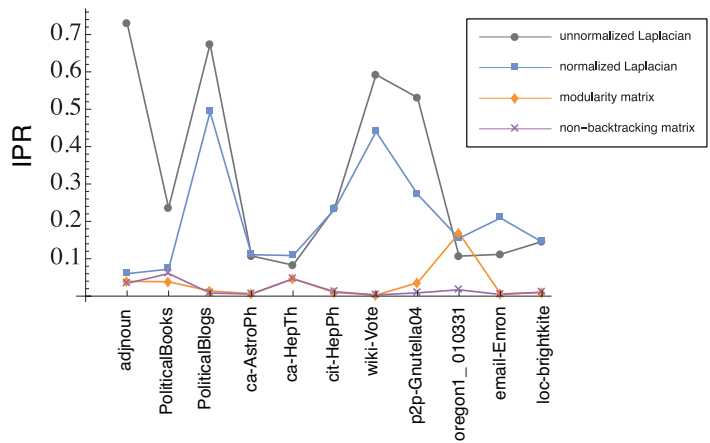}
\caption{ 
(Color online) 
The values of largest IPR among $6$ eigenvectors with the largest or smallest eigenvalues in $11$ real networks. 
First $3$ networks, 
\textit{adjnoun} \cite{Newman2006PRE}, 
\textit{Political books} \cite{Newman2006}, 
\textit{Political blogs} \cite{polblogs} 
are the datasets distributed by Mark Newman \cite{NewmanDataset}. 
Other networks, 
\textit{ca-Astro} \cite{Leskovec2007}, 
\textit{Phca-HepTh} \cite{Leskovec2007}, 
\textit{cit-HepPh} \cite{cit-HepPh1,cit-HepPh2}, 
\textit{wiki-Vote} \cite{wiki-Vote1,wiki-Vote2}, 
\textit{p2p-Gnutella04} \cite{Leskovec2007,Ripeanu02mappingthe}, 
\textit{oregon1\_010331} \cite{Leskovec2005}, 
\textit{email-Enron} \cite{Leskovec2009,KlimtY04}, and 
\textit{loc-brightkite} \cite{Cho2011} are of SNAP \cite{SNAP}; 
the datasets are labeled as listed in \cite{SNAP}. 
}
\label{realIPRs}
\end{figure}

In Fig.~\ref{realIPRs}, we plotted the largest IPR of the $6$ smallest or largest eigenvectors of each matrix. 
We plotted the small eigenvalues for the unnormalized and normalized Laplacians, $L$ and $\mathcal{L}$, 
and the large (real) eigenvalues for the modularity and non-backtracking matrices, $M$ and $B$. 
Note that the non-backtracking matrix may have less than $6$ real eigenvalues. 
In many networks, the IPRs of unnormalized and normalized Laplacian, $L$ and $\mathcal{L}$, can be very large, while those of the modularity matrix $M$ tend to be very small. 
In particular, the non-backtracking matrix $B$ never exhibits a high IPR. 
It is interesting to note that the modularity matrix $M$ is robust to localization as the non-backtracking matrix $B$ very often. Concerning the degree of localization, in practice, it can be inferred that localization can rarely be problematic when $M$ is used. 

\section{Localized eigenvectors of the non-backtracking matrix}\label{LocalizedNB}
We have observed that the eigenvectors of non-backtracking matrix are quite robust against localization. 
It might be expected that the non-backtracking matrix exhibits high IPRs when many short loops exist. 
However, it is hard to obtain high IPRs by random generation of triangles (see Appendix). 
A question that needs to be answered is whether it can be proven that the non-backtracking matrix never has a localized eigenvector. 
Although it may be extremely rare in real networks, localized eigenvectors do exist in the non-backtracking matrix.

\begin{figure}[t]
\centering
\includegraphics[width=0.4\columnwidth]{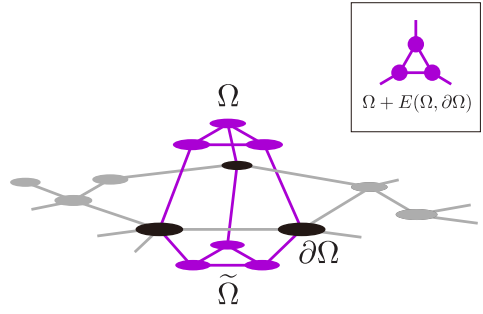}
\caption{ 
(Color online) 
An example of motif doubling with a clique of $3$ nodes. 
}
\label{MotifDoubling}
\end{figure}

It is known that, for the random-walk Laplacian, $I - D^{-1}A$, localized eigenvectors can be constructed owing to the existence of a pair of symmetric subgraphs. 
The construction of a graph with such a local symmetry is called \textit{motif doubling} \cite{Banerjee2008}. 
The procedure of motif doubling is as follows. 
Let $\Omega$ be a set of nodes of a subgraph (or a motif) in a graph $G(V,E)$, where $V$ and $E$ denote the sets of nodes and links of the entire graph, respectively. 
We denote the neighbors of $\Omega$ by $\partial \Omega$. 
We assume that $\Omega$ is only a small part of the entire graph, i.e., $|\Omega| \ll |V|$. 
We build a copy of the induced subgraph of $\Omega$, whose node set is denoted as $\widetilde{\Omega}$. For each node $\tilde{i}_{\alpha} \in \widetilde{\Omega}$, there exists a corresponding node $i_{\alpha} \in \Omega$. 
We then connect this copied subgraph to the nodes in $\partial \Omega$ in the same manner as the nodes in $\Omega$ do. That is, if $i_{\alpha}$ is connected to some nodes in $\partial \Omega$, $\tilde{i}_{\alpha}$ is also connected to those particular nodes (see Fig.~\ref{MotifDoubling} for example). 
In \cite{Banerjee2008}, it is shown that the entire graph that is generated by the motif doubling has a right eigenvector of the random-walk Laplacian whose component is concentrated within $\Omega \cup \widetilde{\Omega}$. 
For its explicit form, see \cite{Banerjee2008} or refer to our derivation for the non-backtracking matrix below. 
Although \cite{Banerjee2008} considers the procedure of doubling as explained above, hereafter, we regard that this locally symmetric subgraph naturally exists in the graph that we consider.

We can apply the motif doubling to obtain the eigenvectors of the non-backtracking matrix $B$. 
Its eigenvalue equation is $A \ket{v} = \mu \ket{v} +  \mu^{-1}(D - I) \ket{v}$, where $\mu$ is the eigenvalue. 
We restrict ourselves to the case where the induced subgraph of $\Omega$ is a regular graph and every node in $\Omega$ is connected to the nodes in $V \backslash(\Omega\cup\widetilde{\Omega})$ by a single link (see Fig.~\ref{MotifDoubling}), because it makes the eigenvalue equation significantly simple. 
We first consider the eigenvalue equation for $\Omega \cup E(\Omega, \partial \Omega)$, the induced subgraph of $\Omega$, and the links between $\Omega$ and $\partial \Omega$. 
Denoting its corresponding eigenvector as $\ket{v}^{\Omega}$, we have the eigenvalue equation as 
\begin{eqnarray}
\sum_{j \in \Omega} A_{ij} v_{j}^{\Omega} = \left( \mu + \frac{c-1}{\mu} \right) v_{i}^{\Omega}. \label{SubgraphEq}
\end{eqnarray}
Note that, although the nodes in (\ref{SubgraphEq}) are of $\Omega$, the degree of each node is not $c-1$, but $c$, because $E(\Omega, \partial \Omega)$ is included. 
Using $\ket{v}^{\Omega}$, we can construct a localized eigenvector $\ket{v}$ of the entire graph. 
Its components are proportional to 
\begin{eqnarray}
v_{i} = 
\left\{
\begin{array}{ll}
 v^{\Omega}_{i} \hspace{20pt} & \textrm{for } i = i_{\alpha} \in \Omega, \\
 -v^{\Omega}_{i} \hspace{20pt} &\textrm{for } i = \tilde{i}_{\alpha} \in \widetilde{\Omega}, \\
 0 \hspace{20pt} &\textrm{otherwise}. \label{VectorElements}
\end{array}
\right.
\end{eqnarray}
It can be easily confirmed that this is indeed an eigenvector of the entire graph. 
Though the nodes in $V\backslash\partial \Omega$ trivially satisfy the eigenvalue equation with $\mu$, for a node $g \in \partial \Omega$, using (\ref{VectorElements}), we have 
\begin{eqnarray}
\sum_{i \in V} A_{gi} v_{i} 
&= \sum_{i \in \Omega} A_{gi} v_{i} + \sum_{i \in \widetilde{\Omega}} A_{gi} v_{i} \nonumber\\
&= \left( \mu + \frac{c-1}{\mu} \right)^{-1} \sum_{i \in \Omega} A_{gi} \sum_{j \in \Omega} A_{ij} v_{j} 
+ \left( \mu + \frac{c-1}{\mu} \right)^{-1} \sum_{i \in \widetilde{\Omega}} A_{gi} \sum_{j \in \widetilde{\Omega}} A_{ij} v_{j} \nonumber\\
&= 0. 
\end{eqnarray} 
Therefore, the vector of (\ref{VectorElements}) satisfies the eigenvalue equation of the entire graph, yielding 
\begin{eqnarray}
\mathrm{IPR} = \frac{1}{2|\Omega|}. 
\end{eqnarray}

In the following sections, we demonstrate the emergence of the localized eigenvector (\ref{VectorElements}) using synthetic graphs and show that its eigenvalue can indeed be the second-largest one.

\section{A pair of cliques attached to a stochastic block model}\label{MotifDoublingSBM}
\begin{figure}[!t]
\centering
\includegraphics[width=0.9\columnwidth]{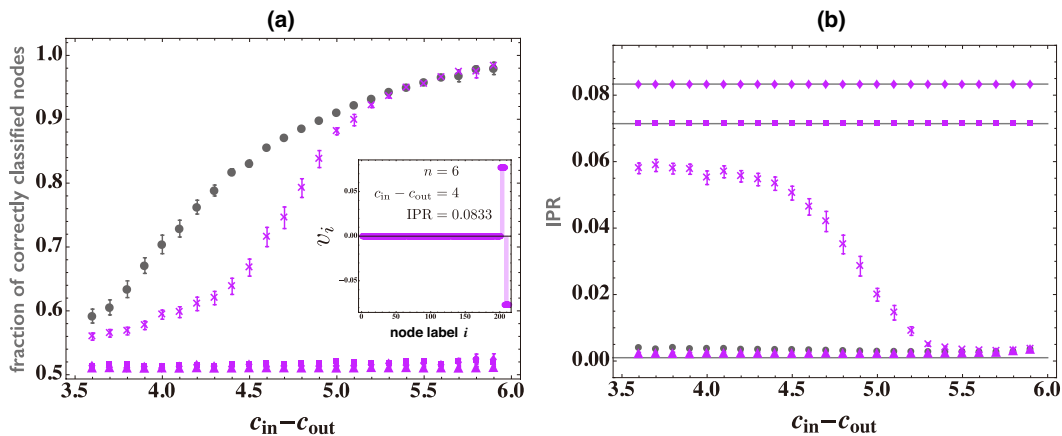}
\caption{ 
(Color online) 
(a) Fraction of correctly classified nodes and (b) the IPR as a function of the strength of block structure $c_{\mathrm{in}} - c_{\mathrm{out}}$ for the sparse stochastic block models with a pair of cliques. 
As the stochastic block model, we set $N=1000$ for two equal size modules and the average degree $\overline{c} = 3$; the detectability threshold of the non-backtracking matrix is at $c_{\mathrm{in}} - c_{\mathrm{out}} = 2\sqrt{3} \sim 3.46$. 
In both plots, circles represents the result of the standard stochastic block model, while the crosses, triangles, diamonds, and squares represent the cases where the clique pair with $n=4,5,6,7$ are attached, respectively. 
The plots show the average over $100$ samples with error bars. 
In (b), the solid lines indicate the values of IPRs equal to $1/12$, $1/14$, and $1/N$. 
The inset of (a) represents the eigenvector elements of the second-largest eigenvalues of a single realization. 
For this particular case, we set $N=200$ and $c_{\mathrm{in}} - c_{\mathrm{out}} = 4$, $n=6$. 
The first $n$ nodes indicate the randomly chosen nodes in $\partial \Omega$ and the last $2n$ nodes indicate the nodes in $\Omega$ and $\widetilde{\Omega}$. 
}
\label{localizedNBSBM}
\end{figure}

As a first example, let us consider a clique of $n$ nodes as $\Omega$ and a sparse stochastic block model as the base graph $V \backslash(\Omega\cup\widetilde{\Omega})$. The neighboring nodes of the clique $\partial \Omega$ are chosen randomly. 
The stochastic block model is constructed as follows. 
We denote the total number of nodes $N$ and each node has its planted module assignment $r$, i.e., for the two planted module case, $r = 1,2$. 
Every node pair is connected at random on the basis of their module assignments. 
Here, we consider a fundamental stochastic block model of two modules with an assortative structure; 
a node pair is connected with probability $p_{\mathrm{in}}$ if they belong to the same module, and with probability $p_{\mathrm{out}}$ ($p_{\mathrm{out}} < p_{\mathrm{in}}$) if they belong to different modules. 
That is, the nodes in the same module are more densely connected than the node in other modules. 
In order to impose the constraint that the graph is sparse, we set both $p_{\mathrm{in}}$ and $p_{\mathrm{out}}$ to be of $O(1/N)$, i.e., the average degree remains constant as we increase the total number of nodes $N$. 
Following the notation in the literature, let us introduce variables $c_{\mathrm{in}} = p_{\mathrm{in}} N = O(1)$ and $c_{\mathrm{out}} = p_{\mathrm{out}} N = O(1)$. 
Because each node pair is connected independently randomly, the graph has a Poisson degree distribution; further, because the graph is sparse, it is locally tree-like. 
As the block structure gets weaker, i.e., $c_{\mathrm{in}} - c_{\mathrm{out}}$ is smaller, the result of the spectral method loses the correlation with the planted solution. 
The value at which they become completely uncorrelated is called the detectability threshold \cite{Decelle2011,Decelle2011a,Krzakala2013,Kawamoto2015}; the region above this threshold is the detectable phase, whereas region below it is the undetectable phase. 

We now solve for (\ref{SubgraphEq}). In the present case, the degree of each node is $n$.  
If we consider the factor $\mu + (n-1)/\mu$ as another constant $\lambda$, it just represents an eigenvalue equation for the adjacency matrix of the induced subgraph of $\Omega$. The eigenvalues of a clique are $-1$ and $n-1$. 
Because we are interested in an eigenvector that is out of the spectral band, we solve for the solution with a real eigenvalue. 
For $\lambda = n-1$, we have 
\begin{eqnarray}
&\mu^{2} - (n-1)\mu + n-1 = 0, \label{mu-equation}\\
&\mu = \frac{1}{2} \left( n - 1 \pm \sqrt{(n-1)(n-5)} \right). \label{mu-solution}
\end{eqnarray}
Therefore, the eigenvalue can be real for $n\ge5$. 
For the solution with $\lambda = -1$, we always have a complex eigenvalue for $\mu$. 
Note that, a real eigenvalue does not always indicate that it is out of the spectral band. 
Moreover, it may not be the second-largest eigenvalue. 
As we demonstrate below, however, it can actually be the eigenvector with the second-largest eigenvalue. 

For the stochastic block model, it is known that the spectral method of the non-backtracking matrix $B$ is free of localization \cite{Krzakala2013}. 
On the other hand, when the clique pair is attached as described in motif doubling, as shown in Fig.~\ref{localizedNBSBM}(a), the eigenvector of the second-largest eigenvalue can be uncorrelated to the planted partition even in the region where the block structure is fairly strong. 
According to (\ref{mu-solution}), we have $\mu \simeq 3.618$ for $n=6$ and it is out of the edge of the spectral band $\sqrt{\overline{c}} \simeq 1.732$ of the standard stochastic block model. 
Figure \ref{localizedNBSBM}(b) indicates that such spectral partitions exhibit high IPRs, which coincide with $(2n)^{-1}$ for $n > 5$. 
Therefore, we conclude that the eigenvector we considered in the previous section can indeed be the one with the second-largest eigenvalue. 
Although the eigenvector of $n=5$ is also uncorrelated, it is extended. 
Interestingly, Fig.~\ref{localizedNBSBM}(b) shows that the eigenvectors with high IPRs gradually appear when $n=4$ as the block structure gets weaker. Accordingly, the decrease of correlation with the planted partition hastens.  
However, we note that this eigenvector has nonzero weights for the components corresponding to the nodes outside the cliques and is different from the eigenvector in (\ref{VectorElements}).

\section{A pair of regular graphs attached to a regular graph}
\begin{figure}[!t]
\centering
\includegraphics[width=\columnwidth]{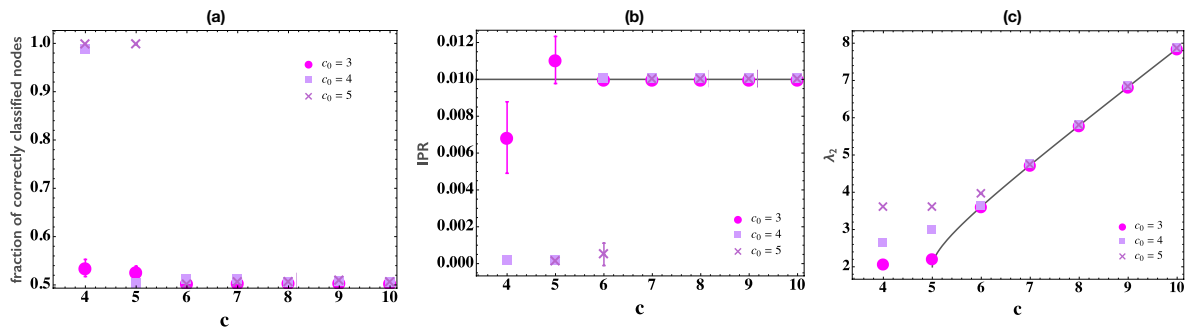}
\caption{ 
(Color online) 
(a) Fraction of correctly classified nodes, (b) IPR, and (c) second-largest eigenvalue as a function of the degree $c$ in $\Omega$ for various values of $c_{0}$. 
As the stochastic block model, we set $N=10000$ for two equal size modules and $c_{\mathrm{in}} - c_{\mathrm{out}} = 2 c_{0} - 1$, i.e., the base graph has a very clear block structure. 
We set $|\Omega| = 50$, so that the value of the IPR for the eigenvector of (\ref{VectorElements}) is $0.01$, which is indicated as a solid line in (b).
The plots show the average over $20$ samples with error bars. 
In (c), solid line indicates the eigenvalue $\mu$ in (\ref{mu-solution-Regular}). 
}
\label{localizedNBTregular}
\end{figure}

To show that setting a clique for $\Omega$ is not essential, we demonstrate another example. 
We consider a random $(c-1)$-regular graph as the induced subgraph of $\Omega$ and a random $c_{0}$-regular graph with block structure as the base graph $V \backslash(\Omega\cup\widetilde{\Omega})$. 
As mentioned in Sec.~\ref{LocalizedNB}, each node in $\Omega$ is connected to a node in the base graph by a single link; 
i.e., the degree of a node in $\Omega$ is $c$. 
The eigenvalue problem of the present case contains the previous example as a special case. 
The leading eigenvalue for $A^{\Omega}$ is $c-1$ and the eigenvalue $\mu$ of the non-backtracking matrix is 
\begin{eqnarray}
&\mu = \frac{1}{2} \left( c - 1 \pm \sqrt{(c-1)(c-5)} \right). \label{mu-solution-Regular}
\end{eqnarray}

Figure \ref{localizedNBTregular} shows the results of numerical experiments and the estimates for the localized eigenvector of (\ref{VectorElements}). 
They show that, when $c \ge 6$ and $c \ge c_{0}+1$ are satisfied, we can obtain the localized eigenvector of (\ref{VectorElements}) as the one with the second-smallest eigenvalue, which is analogous to the previous example.

\section{Discussion}\label{Discussion}
For commonly used matrices for graph partitioning and the non-backtracking matrix, we conducted comparative analyses of the eigenvector localization for several examples of real networks. 
The non-backtracking matrix did not show high values of IPR in any of these analyses. 
We also observed that the modularity matrix $M$ is as robust against localization as the non-backtracking matrix $B$ in many cases. 
Although the localized eigenvector of the non-backtracking matrix seems rare in practice, we constructed an example that exhibits localization. 

The eigenvector of (\ref{VectorElements}) is particularly interesting because it is strongly localized; 
the eigenvector elements are exactly zero outside the pair of motifs due to the symmetry. 
In general, the elements of a localized eigenvector far from the center of localization do not need to be exactly zero. 
As we observed in Sec.~\ref{MotifDoublingSBM}, the eigenvector of (\ref{VectorElements}) is not the only one that exhibits a high IPR. 
However, the forms of eigenvectors of other types may depend on global quantities, e.g., size of the entire graph. 
Thus, it is more difficult to evaluate their IPRs. 
Moreover, having a high IPR is not sufficient to conclude that the eigenvector is uninformative with respect to global structure. 

Even when we have the localized eigenvector of (\ref{VectorElements}), it may be possible to interpret that the spectral method correctly identified the symmetric pair of subgraphs as a characteristic structure. 
However, this eigenvector obtained through the spectral method is not an indicator that we expect to obtain in graph partitioning; the eigenvector of (\ref{VectorElements}) does not convey any information about the module assignment of the nodes outside of $\Omega \cup \partial \Omega$. 
Therefore, considering that, the information regarding global structure is lost. 

The existence of localized eigenvectors out of the spectral band implies that the standard procedure that utilizes $q$ leading eigenvectors for $q$ modules may fail in practical situations. 
Note, however, that it does not necessarily mean that the detection is impossible; by searching more eigenvectors, one might find informative eigenvectors. 
To the best of our knowledge, the condition of the possible case is an open question for any matrix.


\section*{Acknowledgements}
The author would like to thank Yoshiyuki Kabashima for useful comments. 
This work was supported by
JSPS KAKENHI No. 26011023 and
the JSPS Core-to-Core Program ``Non-equilibrium dynamics of soft matter and information.''

\appendix

\section{Degree of localization in the stochastic block model with triangles}\label{ClusteredSBM}
The dynamic interpretation of the non-backtracking matrix $B$ is that, when we regard $B$ as a transition matrix, it prevents a walker from traveling back on the link it has just traveled. 
On the other hand, localization can be understood as the phenomenon that the walker visit a few characteristic nodes very frequently in the case of the adjacency matrix. 
Hence, it makes sense that the non-backtracking matrix avoids localization in sparse random graphs, because short loops are rare. 
This motivated us to analyze how it performs for the graphs with many short loops. 
Because there are many short loops in real networks even though the graph is globally sparse, this setting is also interesting for other matrices. 

We investigate whether localized eigenvectors emerges in random graphs with many triangles. 
Our goal is to verify if a larger number of short loops tend to destroy or enhance the localized eigenvector. 
To observe this tendency, let us consider perturbing a sparse random graph so that the number of triangles is increased. 
As the base random graph, we again use a sparse stochastic block model with two planted modules (see the description in the main text for the details of the stochastic block model). 
The simplest algorithm to increase the number of triangles in a graph is to add some links. 
However, such algorithm changes the degree distribution simultaneously and makes it difficult for us to study the effect solely by short loops. 
For the same reason, we also wish to avoid increasing the degree correlation as much as we can. 
To this end, we employ the algorithm proposed in \cite{Bansal2009}. 
In the algorithm of \cite{Bansal2009}, links are randomly rewired so that the number of triangles increases, while keeping the degree sequence of the original graph. 
Although there are many centralities that measure the degree of clustering, because we just need to know the degree of perturbation, we simply count the number of triangles increased, i.e., $\Delta \tau$ compared with the base graph and normalize it by $N$. 
Although the algorithm allows the graph to be disconnected, we keep the graph connected. 

\begin{figure}[!h]
\centering
\includegraphics[width=0.8\columnwidth]{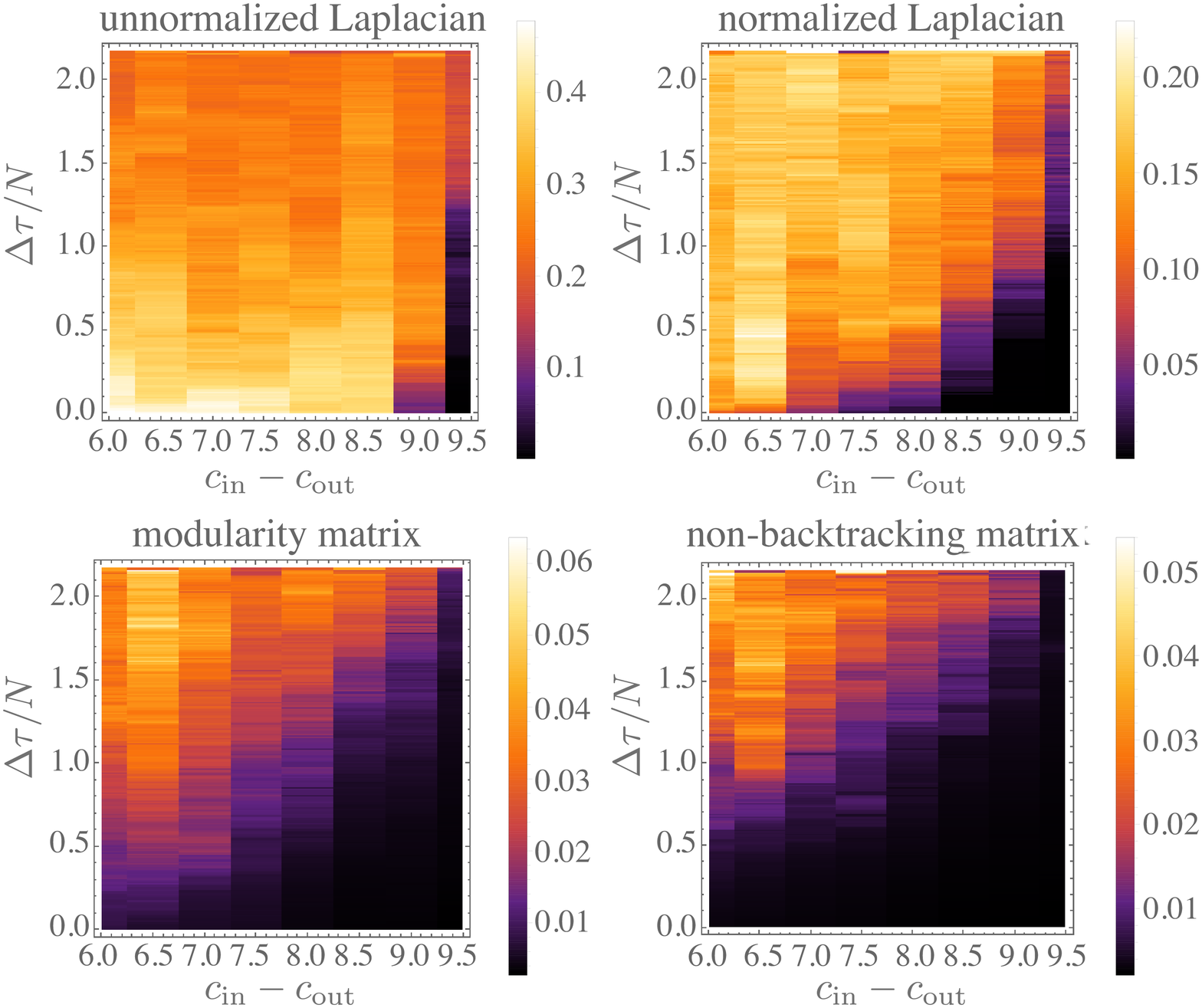}
\caption{ 
(Color online) 
The density plot of the IPR in stochastic block models with controlled number of triangles. 
The horizontal axis represents the strength of block structure $c_{\mathrm{in}} - c_{\mathrm{out}}$; the larger is this value, the stronger is the block structure. 
The vertical axis represents the number of triangles increased, i.e., $\Delta \tau$ by the rewiring algorithm, divided by the total number of nodes $N$. 
For the unnormalized and normalized Laplacians, $L$ and $\mathcal{L}$, the IPRs of eigenvectors with the second-smallest eigenvalues are plotted, whereas the IPRs of the eigenvectors with the largest and second-largest eigenvalues are plotted for the modularity matrix $M$ and the non-backtracking matrix $B$, respectively. 
The plots show the average taken over $20$ samples. 
As the base stochastic block model, we set two equal size modules, where the sum of their nodes is $N = 1000$, and the average degree $\overline{c} = 5$.
}
\label{SBMclosure}
\end{figure}

The results for the benchmark graphs are shown in Fig.~\ref{SBMclosure}. 
We plotted the values of the IPR of eigenvectors of the unnormalized Laplacian $L$, the normalized Laplacian $\mathcal{L}$, the modularity matrix $M$, and the non-backtracking matrix $B$, respectively. 
For the Laplacians, $L$ and $\mathcal{L}$, short loops considerably enhance the localization, although for the unnormalized Laplacian $L$, the IPR is slightly suppressed as $\Delta \tau$ increases in the region where the localized eigenvectors already exist before rewiring. 
As we observed in the analysis of real networks, the modularity matrix $M$ is significantly robust against localization. Although localized eigenvectors may emerge in a larger graph with more triangles, in the present case, the IPR is considerably small in all regions, and therefore we conclude that the localization problem is expected to be rare in the modularity matrix. 
Despite many short loops, the non-backtracking matrix $B$ does not exhibit high IPRs. 
However, we may have larger values of IPRs in a graph with an even stronger clustering structure. 

\section*{References}
\bibliographystyle{iopart-num}
\bibliography{iopart-num}

\end{document}